\documentclass[11pt]{article}

\usepackage[preprint]{acl}

\usepackage{times}
\usepackage{latexsym}
\usepackage{amsmath}
\usepackage{enumitem}
\usepackage{enumerate}

\usepackage{graphicx}
\usepackage{booktabs}
\usepackage{multirow}
\usepackage{array}
\usepackage[table]{xcolor}
\usepackage{algorithm}
\usepackage{algpseudocode}
\usepackage[most]{tcolorbox}
\usepackage{cuted}
\definecolor{criticblue}{RGB}{230, 240, 250}
\definecolor{promptframe}{RGB}{70, 110, 170}
\definecolor{promptbg}{RGB}{245, 248, 253}
\definecolor{slotcolor}{RGB}{180, 70, 30}

\newcommand{\promptslot}[1]{{\color{slotcolor}\rmfamily\itshape\{#1\}}}
\newcommand{\prompttag}[1]{{\color{promptframe}\textless #1\textgreater}}
\newcommand{\promptetag}[1]{{\color{promptframe}\textless /#1\textgreater}}

\newtcolorbox{promptbox}[1]{
  enhanced,
  breakable,
  colback=promptbg,
  colframe=promptframe,
  boxrule=0.5pt,
  arc=2pt,
  left=8pt, right=8pt, top=6pt, bottom=6pt,
  fonttitle=\bfseries\sffamily\small,
  coltitle=white,
  attach boxed title to top left={xshift=8pt, yshift=-2mm},
  boxed title style={colback=promptframe, arc=2pt, boxrule=0pt},
  title={#1},
  fontupper=\small\ttfamily,
  before skip=10pt, after skip=10pt,
}

\usepackage[T1]{fontenc}

\usepackage[utf8]{inputenc}

\usepackage{microtype}

\usepackage{inconsolata}

\usepackage{graphicx}

\title{Critic-R: Improving Agentic Search using Instruction-tuned Retrievers with Natural Language Introspective Feedback}

\author{Md Zarif Ul Alam, Alireza Salemi, Hamed Zamani \\
  Center for Intelligent Information Retrieval \\
  University of Massachusetts Amherst \\
  United States \\
  \texttt{\{zarifalam,asalemi,zamani\}@cs.umass.edu}
  }

\begin{document}
\maketitle
\begin{abstract}

Agentic search systems iteratively interact with retrieval models to answer complex queries. Despite substantial progress, optimizing retrievers for agentic search remains challenging, often requiring heavy co-training or gold-standard annotations that limit real-world applicability. We propose \textbf{Critic-R}, a framework that explicitly closes the feedback loop between the reasoning agent and the retrieval model during both inference and training. Critic-R introduces a critic model that evaluates the agent's introspective reasoning trace after consuming retrieved evidence to determine whether the retrieved context sufficiently supports the next reasoning step. Critic-R has two complementary mechanisms: \textbf{Critic-R-Zero}, an inference-time query refinement loop that iteratively rewrites queries and retrieval instructions, and \textbf{Critic-Embed}, an optimization approach for retrieval models that leverages successful and failed refinement trajectories as automatic supervision without requiring manual relevance annotation. We evaluate Critic-R on {HotpotQA}, {2WikiMultihopQA}, {MuSiQue}, and {Bamboogle}. Results show that Critic-R significantly improves both retrieval quality and downstream answer accuracy.

\end{abstract}

\section{Introduction}

Retrieval-Augmented Generation (RAG) extends Large Language Models (LLMs) with non-parametric access to external corpora and has become a standard framework for knowledge-intensive tasks \citep{RAG:2020,REML:2022}. Early RAG systems primarily relied on single-turn pre-generation retrieval. However, this setting is often insufficient for complex queries that require decomposition or information synthesis from multiple sources. As a result, recent work has shifted toward \textit{agentic search}, in which reasoning models interleave internal deliberation with iterative retrieval actions over multiple steps. Recent advances have shown that reinforcement learning (RL) can be used to optimize these agents directly from task rewards, leading to search-aware reasoning systems such as Search-R1 \citep{jin2025search} and DeepResearcher \citep{zheng2025deepresearcher}.

Most existing agentic search approaches primarily optimize the reasoning agent while treating the retrieval model as a frozen black-box component. This design implicitly assumes that a sufficiently capable reasoning model can compensate for retrieval failures through improved query reformulation alone. This paper challenges this assumption by arguing that sub-optimal retrieval can be a bottleneck in agentic search performance. Recent studies such as Agentic-R \citep{liu2026agentic} and CoSearch \citep{zeng2026cosearch} attempt to address this issue by jointly optimizing retrievers and reasoning agents. In practice, however, these methods are difficult to apply in settings where the reasoning model cannot be further trained, the retriever is externally provided, or gold-passage supervision is unavailable.

\begin{figure*}[t]
\centering
\includegraphics[width=\textwidth]{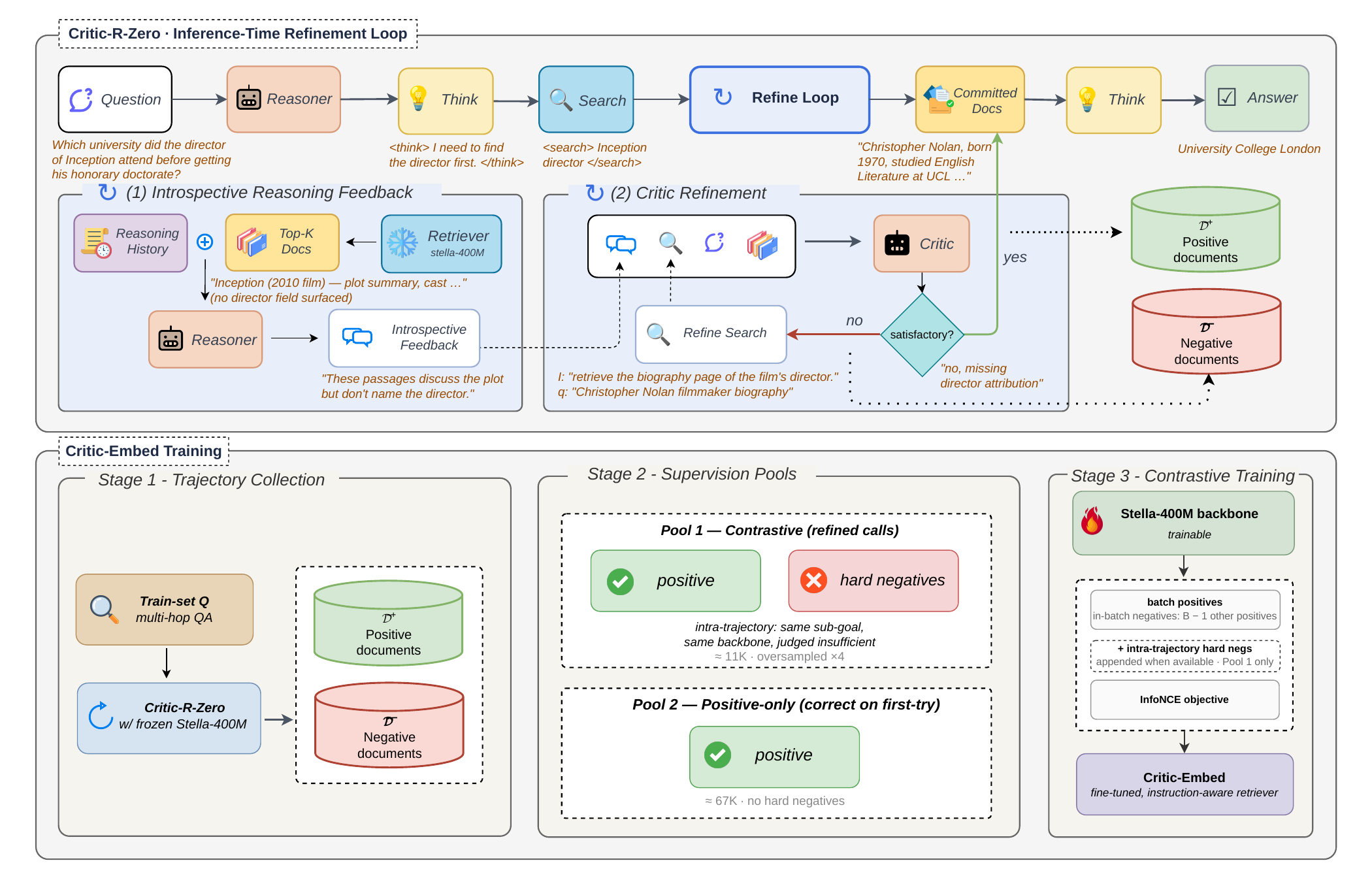}
\caption{Critic-R Overview.}
\label{fig:overview}
\vspace{-0.4cm}
\end{figure*}

To address this, we propose \textbf{Critic-R}, a framework that closes the feedback loop between the reasoning agent and retriever during both inference and training time. Instead of blindly accepting the retrieved documents provided by the retriever, Critic-R enables the agent to assess whether they satisfy its current information requirements before proceeding to subsequent retrieval or reasoning steps. We employ a separate critic model for this purpose for two reasons. First, it allows the framework to be applied to arbitrary reasoning agents without requiring built-in self-criticism or modifications to the underlying model. Second, in long multi-step trajectories, accumulated context and reasoning noise can cause the reasoning model to become overconfident or less sensitive to retrieval failures \cite{jin2025reasoninghurtinductiveabilities}, motivating the use of a dedicated evaluation component. To achieve this, the critic analyzes the agent's introspective reasoning trace---namely, the reasoning generated immediately after consuming the retrieved evidence---to determine whether the retrieved context is sufficient for the next reasoning step. This design uses the observation that the agent often explicitly indicates whether the retrieved documents contain the information required to continue reasoning.

This verification signal enables two complementary mechanisms: (1) \textbf{Critic-R-Zero (Inference-Time Scaling):} An iterative reasoning-evaluation loop that operates entirely at inference time without additional gradient updates. When the critic determines that the retrieved evidence is insufficient, it rewrites the retrieval query and instruction for another retrieval attempt. This process continues until the agent is satisfied with the retrieved evidence or a predefined refinement budget is exhausted. Importantly, the reasoning agent itself remains unchanged and only interacts with retrieved documents, while the refinement process is handled externally. This mechanism dynamically allocates additional inference-time computation to recover from retrieval failures. (2) \textbf{Critic-Embed (Retriever Fine-Tuning):} To amortize the computational overhead introduced by iterative refinement, we leverage the execution trajectories generated by Critic-R-Zero as a source of automatic supervision for retrieval model training. Documents that satisfy the reasoning agent are treated as positive examples, while documents rejected during unsuccessful refinement attempts are treated as hard intra-trajectory negatives. Using this intra-trajectory contrastive learning, we fine-tune the retrieval model without requiring relevance information (i.e., gold passages). These two components are complementary and can be combined within a unified system, where a trained retriever can benefit from the inference-time refinement capabilities.

We evaluate Critic-R on several challenging multi-hop question answering benchmarks, including \textsc{HotpotQA}~\cite{yang2018hotpotqa}, \textsc{2WikiMultihopQA}~\cite{ho2020constructing}, \textsc{MuSiQue}~\cite{trivedi2022musique}, and \textsc{Bamboogle}~\cite{press2023measuring}. Our results demonstrate that explicitly modeling retrieval quality within the agentic reasoning loop leads to substantial improvements in downstream answer accuracy. First, we show that the inference-time refinement mechanism, \textbf{Critic-R-Zero}, significantly alleviates retrieval failures across reasoning agents of different scales by iteratively evaluating retrieved evidence and refining retrieval queries, yielding an overall relative improvement of 12.4\%. We further demonstrate that the fine-tuned dense retriever, \textbf{Critic-Embed}, consistently outperforms both off-the-shelf retrievers and prior co-trained baselines, achieving up to an overall 7.5\% relative improvement. Finally, combining the trained retriever with the inference-time refinement loop---i.e., integrating \textbf{Critic-R-Zero} and \textbf{Critic-Embed} into a unified system called \textbf{Critic-R}---yields the strongest overall performance, achieving a 10.9\% relative improvement overall. To support future research on this topic, we release our code, data, and trained models.\footnote{Available at: \url{https://github.com/zarif98sjs/Critic-R}}

\section{Related Work}

\paragraph{Retrieval-Augmented Generation and Agentic Search.}
Retrieval-Augmented Generation (RAG) extends Large Language Models with non-parametric access to external corpora and has become the standard recipe for knowledge-intensive QA~\cite{asai2023retrieval, gao2023retrieval, ram2023context}. Early systems issued a single query at the start of generation, which is ill-suited to multi-hop questions whose information needs only become apparent partway through reasoning~\cite{shao2023enhancing, trivedi2023interleaving}. Two lines of work have addressed this. The first uses prompting to interleave reasoning with retrieval, exemplified by IRCoT~\cite{trivedi2023interleaving} and ReAct~\cite{yao2022react}. The second teaches models when and how to retrieve through supervised fine-tuning, including Self-RAG~\cite{asai2023self} and Toolformer~\cite{schick2023toolformer}. More recently, reinforcement learning has enabled agents to acquire multi-turn search policies directly from task-outcome rewards: Search-R1~\cite{jin2025search}, R1-Searcher~\cite{song2025r1}, ReSearch~\cite{chen2026learning}, and DeepResearcher~\cite{zheng2025deepresearcher} all train an LLM to alternate \texttt{<think>}, \texttt{<search>}, and \texttt{<answer>} turns, producing a search-aware reasoner. Our work builds directly on this paradigm. We use Search-R1 as our reasoning agent, but is orthogonal to its training objective: rather than modifying how the agent is trained, Critic-R intervenes at inference time to inspect and repair the agent's individual retrieval calls. The gains from improving \emph{how the agent interacts with retrieval} are largely orthogonal to gains from improving the agent itself.

\paragraph{Retrieval optimization for Agents.}
A complementary line of work also rejects the frozen-retriever assumption, but addresses it through additional \emph{training} of the retrieval side. REPLUG~\cite{shi2024replug} and Atlas~\cite{izacard2023atlas} optimize the retriever using generator likelihood as a signal; later approaches use task-level metrics or LLM-judged passage utility~\cite{xu2025training, zamani2024stochastic, zhang2025llm}. Two recent works extend this idea explicitly to the agentic-search setting and make the retrieval bottleneck their central claim. Agentic-R~\cite{liu2026agentic} trains a retriever tailored for multi-turn search by jointly modeling local query-passage relevance and global answer correctness, and iteratively co-optimizes the retriever with the agent. CoSearch~\cite{zeng2026cosearch} quantifies the bottleneck directly through an oracle-retrieval experiment. They show double-digit relative F1 gains when correct documents are guaranteed to appear and jointly train a generative reranker alongside the reasoning agent with GRPO, using a composite reward over ranking quality and final answer correctness. We share the diagnosis with these works but split the problem along a different seam. The first is Critic-R-Zero, a purely inference-time loop that requires no gradient updates anywhere: a separate critic model inspects each retrieval, judges whether the returned context is sufficient based on the reasoning agent's feedback, and rewrites the search instruction and query when it is not. Critic-R-Zero treats the underlying retriever as a fixed black box and is therefore composable with any retriever, including those produced by Agentic-R or CoSearch. The second part, Critic-Embed, is a retriever trained on the trajectories that Critic-R-Zero collects on the train splits of two QA datasets, turning the critic's free-form feedback into supervision without ever requiring gold passage annotations. The full system, Critic-R, is Critic-R-Zero running on top of Critic-Embed. This decomposition allows us to address both lines of prior work: for frozen-retriever agentic search, we introduce Critic-R-Zero; and for retriever-training approaches, we provide a training recipe supervised entirely by the inference loop's own feedback.

\paragraph{Inference-Time Scaling for Reasoning.}
Parallel to the work above, recent work shows that allocating more compute at inference through longer chain-of-thought, self-consistency, or process supervision~\cite{wei2022chain, wang2022self, lightman2023let} can rival the gains from scaling model parameters. OpenAI's o1-style models and subsequent open-source reasoners take this further by training models to spend more tokens deliberating before answering. Critic-R-Zero can be viewed as inference-time scaling targeted at the retrieval bottleneck rather than the reasoning trace: each additional refinement attempt and each step-up in critic size is a controlled investment of compute aimed specifically at recovering from a bad retrieval.

\section{The Critic-R Framework}
\label{sec:methodology}

This section presents the approaches that lead to the development of Critic-R. First, we describe Critic-R-Zero, an inference-time critic on retrieval results for query refinement that does not require any additional training. Second, we introduce Critic-Embed by explaining how Critic-R-Zero can be used to optimize retrieval models through a novel intra-trajectory contrastive learning approach. Next, we describe how these two approaches can complement each other to form Critic-R. Last, we describe our implementation details.

\subsection{Critic-R-Zero for Inference-Time Query Refinement}

As shown in Algorithm~\ref{alg:critic-r-zero} and Figure~\ref{fig:overview}, we assume access to a frozen reasoning agent $\mathcal{M}_R$ operating under the ReAct framework \cite{yao2022react}, which is allowed to perform at most $M$ actions to answer a question $Q$ (Line~\ref{line:for_loop_max_iterations}). At each step $i$, the agent $\mathcal{M}_R$ produces a reasoning trace $T_i$ and an action $A_i$ (Line~\ref{line:sample_reasoning_and_action}), which are appended to the overall trajectory (Line~\ref{line:append_to_the_trajectory}). If $A_i$ is a final answer, the trajectory terminates (Line~\ref{line:if_answer_break}). Otherwise, if $A_i$ is a search action, the agent extracts an initial query $q_i^{(1)}$, which is augmented with a default instruction\footnote{The default instruction is: \textit{``Given a query, retrieve relevant passages that answer the query''}.} $I_i^{(1)}$ for the retrieval model $\mathcal{R}$ (Lines~\ref{line:extract_query}--\ref{line:default_instruction}). Entering the search phase, an instruction-aware retrieval model $\mathcal{R}$ returns the top $k$ documents, $D_i^{(t)} = \mathcal{R}(q^{(t)}_i, I^{(t)}_i, k)$ (Line \ref{line:retriever_docs}). Note that these initial retrieved documents frequently fail to provide the necessary evidence, making single-turn retrieval a severe bottleneck for agentic search performance. This dissatisfaction is typically reflected in the model's subsequent reasoning trace. 

To address this limitation, we introduce a speculative refinement loop (Line~\ref{line:refine_query_at_most_k_times}) that allows the agent to recover from an ineffective retrieval. At each refinement step, the retrieved documents $D_i^{(t)}$ are speculatively provided to the reasoner to generate an introspective reasoning trace, without yet committing these documents to the persistent trajectory (Line~\ref{line:sample_next_thinking_and_action_based_on_retrieved_docs}). A separate critic model, denoted as $\mathcal{M}_C$, then evaluates this trace to determine whether the retrieved evidence sufficiently resolves the reasoner's information need (Line~\ref{line:check_if_reasoner_is_satisfied_with_documents}). If the critic produces positive feedback, $\sigma_i^{(t)} = \text{yes}$, the retrieved documents are added to the candidate positive set $\tilde{\mathcal{D}}^{+}$ as useful evidence, and the refinement loop terminates (Lines~\ref{line:add_retrieved_documents_as_positive}--\ref{line:break_refinement_loop}). Otherwise, the documents are assigned to the candidate negative set $\tilde{\mathcal{D}}^{-}$ (Line~\ref{line:add_retrieved_documents_as_negative}). The critic then leverages the reasoner's explicit dissatisfaction in its thinking trace to generate a refined query and retrieval instruction for the next iteration (Line~\ref{line:generate_new_query_and_instruction_for_next_round}). This process repeats for at most $K$ refinement steps, adaptively bridging retrieval failures before the final selected documents are committed to the reasoning history (Line~\ref{line:commit_final_documents_to_history}). The procedure concludes by returning the final extracted answer together with the collected positive and negative document sets (Lines~\ref{line:if_answer_final}--\ref{line:return_final_outputs}). These document sets are used exclusively for retriever training and are not required during inference.

\subsection{Critic-Embed: Intra-Trajectory Contrastive Learning for Training Instruction-Following Dense Retrieval}

While the inference-time refinement loop effectively detects and repairs retrieval failures, repeated interaction with the critic introduces additional computational overhead. To amortize this cost and permanently improve the retrieval backbone without relying on expensive human-annotated gold passages, we leverage the execution trajectories generated by Critic-R-Zero as a source of automatically constructed supervision. Specifically, each refinement trajectory produces a natural training signal: documents that satisfy the reasoner, as verified by the critic, are treated as positive examples ($\mathcal{D}^{+}$), while documents rejected during earlier unsuccessful refinement attempts are treated as hard intra-trajectory negatives ($\mathcal{D}^{-}$). We use the collected supervision signals to fine-tune the retriever, resulting in Critic-Embed, a retriever trained to better align retrieved evidence with the information requirements of the reasoning agent for a given query. To ensure label quality, we retain only trajectories whose final prediction is correct according to the downstream task metric. The contrastive learning loss for each training instance is defined as:
\begin{equation}
\mathcal{L} = - \log \frac{\exp\!\big(\mathrm{sim}(q_i, z_i^+)/\tau\big)}{\sum_{z \in \mathcal{Z}_i} \exp\!\big(\mathrm{sim}(q_i, z)/\tau\big)}
\end{equation}
where $q_i$ is the query embedding, $z_i^+$ is its paired positive document, $\mathcal{Z}_i$ contains the positive, all in-batch negatives, and the query's intra-trajectory hard negatives, $\mathrm{sim}(\cdot,\cdot)$ denotes cosine similarity, and $\tau$ is the temperature.

\subsection{Critic-R: Improving Critic-Embed with Inference-Time Scaling}

The complete Critic-R pipeline pairs the Critic-R-Zero inference loop with the trained Critic-Embed retriever. This composition ensures the agent starts with a highly capable, domain-aligned retrieval that minimizes initial search errors, while still maintaining the inference-time ability to introspect and recover from complex edge-case retrieval failures.

\subsection{Implementation Details}
\label{sec:method:critic-r}

The reasoning agent $\mathcal{M}_R$ is an instruction-tuned LLM operating under the ReAct paradigm~\cite{yao2022react}, alternating \texttt{<think>}, \texttt{<search>}, and \texttt{<answer>} actions, with retrieved documents injected within \texttt{<information>} tags. The critic $\mathcal{M}_C$ is a separate LLM that operates in two sequential modes: a \emph{satisfaction judgment} mode that emits a binary verdict $\sigma_i^{(t)}$ and diagnostic reason $r_i^{(t)}$ conditioned on $(Q, q_i^{(t)}, D_i^{(t)}, T_{i+1}^{(t)})$, and a \emph{query refinement} mode invoked only when $\sigma_i^{(t)}=\texttt{no}$ that rewrites the sub-query and retrieval instruction using $r_i^{(t)}$. Full component descriptions and all prompts are deferred to Appendix~\ref{app:impl-details}.

\begin{algorithm}[t]
\small
\caption{Critic-R-Zero: Inference Loop}
\label{alg:critic-r-zero}
\begin{algorithmic}[1]
\Require Question $Q$, corpus $\mathcal{C}$, reasoner $\mathcal{M}_R$, critic $\mathcal{M}_C$, instruction-aware retriever $\mathcal{R}$, number of retrieved docs $k$, max reasoning iterations $M$, max refinements $K$ \label{line:require_inputs}
\State  $H \gets Q$  \Comment{initialize reasoning history} \label{line:initialize_reasoning_history}
\State  $\tilde{\mathcal{D}}^{+} \gets \emptyset$ \Comment{initialize candidate positive set} \label{line:initialize_candidate_positive_set}
\State $\tilde{\mathcal{D}}^{-} \gets \emptyset$ \Comment{initialize candidate negative set} \label{line:initialize_candidate_negative_set}
\For{$i = 1 \xrightarrow{} M$} \label{line:for_loop_max_iterations}
    \State $(T_i, A_i) \gets \mathcal{M}_R(H)$ \Comment{Sample reasoning \& action} \label{line:sample_reasoning_and_action}
    \State $H \gets H;T_i;A_i$ \Comment{Append to the trajectory} \label{line:append_to_the_trajectory}
    \If{$A_i$ is an \texttt{<answer>} action} \State \textbf{break} \label{line:if_answer_break}
    \Else \label{line:else_not_answer}
    \State $q_i^{(1)} \gets A_i$ \Comment{Extract query} \label{line:extract_query}
    \State $I_i^{(1)} \gets \emptyset$ \Comment{Default instruction} \label{line:default_instruction}
    \EndIf \label{line:end_if_answer}
    \For{$t = 1 \xrightarrow{} K$} \Comment{Refine query at most $K$ times} \label{line:refine_query_at_most_k_times}
        \State $D_i^{(t)} \gets \mathcal{R}(q_i^{(t)}, I_i^{(t)}, k)$ \Comment{Retriever docs} \label{line:retriever_docs}
        \State $T_{i+1}^{(t)}, A_{i+1}^{(t)} \gets \mathcal{M}_R(H;D_i^{(t)})$ \Comment{Sample next thinking \& action based on retrieved docs} \label{line:sample_next_thinking_and_action_based_on_retrieved_docs}
        \State $(\sigma_i^{(t)}, r_i^{(t)}) \gets \mathcal{M}_C(P_{\text{J}}(Q, q_i^{(t)}, D_i^{(t)}, T_{i+1}^{(t)}))$ \Comment{Check if reasoner is satisfied with documents} \label{line:check_if_reasoner_is_satisfied_with_documents}
        \If{$\sigma_i^{(t)} = \texttt{yes}$} \label{line:if_reasoner_satisfied}
            \State $\tilde{\mathcal{D}}^{+} \gets \tilde{\mathcal{D}}^{+} \cup \{(i, q_i^{(t)}, I_i^{(t)}, D_i^{(t)})\}$ \Comment{Add retrieved documents as positive} \label{line:add_retrieved_documents_as_positive}
            \State \textbf{break} \label{line:break_refinement_loop}
        \Else \label{line:else_reasoner_not_satisfied}
            \State $\tilde{\mathcal{D}}^{-} \gets \tilde{\mathcal{D}}^{-} \cup \{(i, q_i^{(t)}, I_i^{(t)}, D_i^{(t)})\}$ \Comment{Add retrieved documents as negative} \label{line:add_retrieved_documents_as_negative}
        \EndIf \label{line:end_if_reasoner_satisfied}
        \State $(q_i^{(t+1)}, I_i^{(t+1)}) \gets \mathcal{M}_C(P_{\text{R}}(Q, q_i^{(t)}, I_i^{(t)}, r_i^{(t)}))$ \Comment{New query and instruction for next refinement round} \label{line:generate_new_query_and_instruction_for_next_round}
    \EndFor \label{line:end_for_refinement}
    \State $H \gets H;D_i^{(t)} \,$ \Comment{commit final documents to history} \label{line:commit_final_documents_to_history}
\EndFor \label{line:end_for_max_iterations}

\If{$A_i$ is an \texttt{<answer>} action} \State $\hat{y} \gets A_i$ \label{line:if_answer_final} \Comment{Extract answer}
    \Else \label{line:not_answer_end}
    \State $\hat{y} \gets \emptyset$ \Comment{No answer generated in $M$ actions} \label{line:no_answer}
    \EndIf
\State \Return $\hat{y}$, $\tilde{\mathcal{D}}^{+}$, $\tilde{\mathcal{D}}^{-}$ \label{line:return_final_outputs}
\Comment{returning final answer, positive docs and negative docs (only for training)}
\end{algorithmic}
\end{algorithm}

\section{Experiments}
\label{sec:experiments}

We structure our evaluation around four questions:

\begin{itemize}[leftmargin=1.2em]
    \item \textbf{RQ1.} Can the retrieval bottleneck in agentic search be mitigated
    \emph{without} modifying the retriever itself, and, how does the gain scale with the critic model's parameters size?
    \vspace{-0.3cm}
    \item \textbf{RQ2.} Do the trajectories that Critic-R-Zero collects contain transferable retrieval supervision i.e., does fine-tuning a retriever with them (Critic-Embed) beat both an off-the-shelf dense retriever and a retriever co-trained end-to-end with the search agent?
    \vspace{-0.3cm}
    \item \textbf{RQ3.} Can combining inference-time query refinement loop and the trained retriever yield further gains?
    \vspace{-0.3cm}
    \item \textbf{RQ4.} Is the agent's introspective feedback $T_{i+1}$ a key source of supervisory signal that Critic-Embed inherits?
\end{itemize}

\subsection{Datasets}
Following previous work \cite{jin2025search}, we evaluate our method on four multi-hop QA datasets requiring synthesizing information across multiple documents: \textbf{HotpotQA}~\cite{yang2018hotpotqa}, \textbf{2WikiMultihopQA}~\cite{ho2020constructing}, \textbf{MuSiQue}~\cite{trivedi2022musique}, and \textbf{Bamboogle}~\cite{press2023measuring}. To assess whether the critic loop also helps when a single-hop retrieval is sufficient, we additionally report results on three general-domain QA datasets: \textbf{NQ}~\cite{kwiatkowski2019natural}, \textbf{TriviaQA}~\cite{joshi2017triviaqa}, and \textbf{PopQA}~\cite{mallen2023not} for Critic-R-Zero experiments. Dataset statistics are reported in Table~\ref{tab:dataset-stats} in Appendix~\ref{app:dataset-stats}. For evaluation, we report \textbf{Exact Match (EM)} and token-level \textbf{F1}, following prior work \cite{jin2025search}.

\begin{table*}[ht]
\centering
\resizebox{1.8\columnwidth}{!}{%
\small
\setlength{\tabcolsep}{4pt}
\renewcommand{\arraystretch}{1.1}
\begin{tabular}{@{}ll cc cc cc cc cc@{}}
\toprule
& & \multicolumn{2}{c}{\textbf{HotpotQA}} & \multicolumn{2}{c}{\textbf{2Wiki}} & \multicolumn{2}{c}{\textbf{Musique}} & \multicolumn{2}{c}{\textbf{Bamboogle}} & \multicolumn{2}{c}{\textbf{Avg.}} \\
\cmidrule(lr){3-4}\cmidrule(lr){5-6}\cmidrule(lr){7-8}\cmidrule(lr){9-10}\cmidrule(lr){11-12}
\textbf{Reasoner} & \textbf{Critic ($\mathcal{M}_C$)} & EM & F1 & EM & F1 & EM & F1 & EM & F1 & EM & F1 \\
\midrule
\multirow{4}{*}{Search-R1 (14B)}
 & \textit{no-critic}     & 0.4149 & 0.5284 & 0.4367 & 0.4946 & 0.1849 & 0.2684 & 0.3520 & 0.4963 & 0.3472 & 0.4470 \\
 & Qwen2.5-14B           & 0.4373 & 0.5539 & 0.4460 & 0.5051 & 0.2069 & 0.2948 & 0.4320 & 0.5510 & 0.3806 & 0.4762 \\
 & Qwen2.5-32B           & \textbf{0.4431} & 0.5586 & 0.4492 & 0.5064 & \textbf{0.2218} & \textbf{0.3110} & 0.4400 & \textbf{0.5668} & 0.3886 & \textbf{0.4857} \\
 & Qwen2.5-72B           & 0.4425 & \textbf{0.5593} & \textbf{0.4580} & \textbf{0.5155} & 0.2127 & 0.3045 & \textbf{0.4480} & 0.5627 & \textbf{0.3903} & 0.4855 \\
\midrule
\multirow{4}{*}{Search-R1 (7B)}
 & \textit{no-critic}     & 0.3589 & 0.4667 & 0.3369 & 0.3963 & 0.1324 & 0.2151 & 0.3600 & 0.4691 & 0.2971 & 0.3868 \\
 & Qwen2.5-14B           & 0.3741 & 0.4815 & 0.3465 & 0.4070 & 0.1398 & 0.2220 & 0.3840 & 0.4984 & 0.3111 & 0.4023 \\
 & Qwen2.5-32B           & 0.3759 & 0.4834 & 0.3533 & 0.4135 & \textbf{0.1568} & \textbf{0.2383} & \textbf{0.4309} & \textbf{0.5351} & \textbf{0.3293} & \textbf{0.4176} \\
 & Qwen2.5-72B           & \textbf{0.3769} & \textbf{0.4865} & \textbf{0.3594} & \textbf{0.4227} & 0.1485 & 0.2360 & 0.3920 & 0.4980 & 0.3192 & 0.4108 \\
\midrule
\multirow{4}{*}{Search-R1 (3B)}
 & \textit{no-critic}     & 0.2813 & 0.3755 & 0.2627 & 0.3169 & 0.0910 & 0.1580 & 0.1920 & 0.2708 & 0.2068 & 0.2803 \\
 & Qwen2.5-14B           & 0.2875 & 0.3841 & 0.2715 & 0.3253 & 0.1005 & 0.1701 & 0.2080 & 0.3095 & 0.2169 & 0.2973 \\
 & Qwen2.5-32B           & 0.2935 & 0.3900 & 0.2749 & 0.3306 & \textbf{0.1030} & \textbf{0.1750} & 0.2177 & 0.3215 & 0.2223 & 0.3043 \\
 & Qwen2.5-72B           & \textbf{0.2987} & \textbf{0.3973} & \textbf{0.2750} & \textbf{0.3323} & 0.0976 & 0.1742 & \textbf{0.2560} & \textbf{0.3571} & \textbf{0.2319} & \textbf{0.3153} \\
\bottomrule
\end{tabular}
}
\vspace{-0.2cm}
\caption{\textbf{Multi-Hop QA --- inference-time scaling along the critic-size axis} for \textbf{Critic-R-Zero} (frozen Stella-400M retriever, top $k\!=\!1$, $K\!=\!2$ refinement attempts). \textbf{Bold} marks the best EM/F1 per (reasoner, dataset, metric).}
\label{tab:its-multihop}
\vspace{-0.3cm}
\end{table*}

\vspace{-0.2cm}
\subsection{Experimental Setup}
We use the optimized checkpoints of {Search-R1}~\cite{jin2025search} for reasoning models, which are instruction-tuned GRPO-trained variants of {Qwen2.5-Instruct\footnote{\url{https://hf.co/collections/PeterJinGo/search-r1-v03}}}. To study scale, we evaluate three sizes: \texttt{SearchR1-Qwen2.5-3B},\texttt{-7B}, and \texttt{-14B}. For the critic, we use frozen instruction-tuned Qwen2.5 with {14B\footnote{\url{https://hf.co/Qwen/Qwen2.5-14B-Instruct}}}, {32B} \footnote{\url{https://hf.co/Qwen/Qwen2.5-32B-Instruct}}, and {72B \footnote{\url{https://hf.co/Qwen/Qwen2.5-72B-Instruct}}} parameters.

The frozen-retriever experiments use a dense retriever based on the \textbf{Stella-400M} embedding model~\cite{zhang2025jasperstelladistillationsota}, with the December 2018 Wikipedia dump~\cite{karpukhin2020dense} indexed as the retrieval corpus for all experiments. The instruction interface is essential to Critic-R-Zero: it is what allows the critic's refined instruction to alter the retrieval behavior without re-indexing. We evaluate three retrieval depths, \textbf{top $k \in \{1, 3, 5\}$}, to measure how the critic's benefit varies as more raw recall is given to the reasoner. 
 
To collect intra-trajectory hard negatives, we run Critic-R-Zero with a Search-R1 (14B) reasoner, a Qwen2.5-72B critic, the frozen Stella-400M backbone, on the \texttt{train} splits of HotpotQA and Musique. The resulting dataset consists of roughly 11K natural contrastive pairs (search calls that underwent at least one refinement, yielding both positives and intra-trajectory hard negatives) and 67K positive-only samples (search calls satisfied on the first attempt).

Critic-Embed is initialized from {Stella-400M}~\cite{zhang2025jasperstelladistillationsota} and fine-tuned with an InfoNCE objective using intra-trajectory hard negatives combined with in-batch negatives, with natural contrastive pairs oversampled relative to positive-only samples. Full training hyperparameters are reported in Appendix~\ref{app:training-details}.

For the full Critic-R experiments, the Stella-400M backbone is replaced by {Critic-Embed}, our fine-tuned retriever. Experiments were run on NVIDIA A100 (80GB) GPUs. We set the maximum number of refinements to \textbf{$K = 2$}, as additional iterations do not yield more improvements.

\paragraph{Baselines.} We compare against two baseline families, depending on the question each table targets:
\begin{itemize}[leftmargin=1.2em]
    \item \textbf{No-critic ablation.} The Critic is removed and Search-R1 is run with default setting. This isolates the contribution of the critic's verdict and refinement (used for RQ1 and RQ3).
    \vspace{-0.2cm}
    \item \textbf{Retriever baselines (Table~\ref{tab:retriever}).} The same Search-R1 reasoner makes a single top $k$ retrieval call against the off-the-shelf Stella-400M backbone and the Agentic-R~\cite{liu2026agentic} retriever, with no critic loop. This isolates the contribution of the retriever itself and lets us compare Critic-Embed against both an untrained dense backbone and a retriever that is co-trained end-to-end with the search agent.
    \vspace{-0.2cm}
\end{itemize}


\begin{table*}[ht]
\centering
\resizebox{1.75\columnwidth}{!}{%
\small
\setlength{\tabcolsep}{4pt}
\renewcommand{\arraystretch}{1.1}
\begin{tabular}{@{}ll cc cc cc cc cc@{}}
\toprule
& & \multicolumn{2}{c}{\textbf{HotpotQA}} & \multicolumn{2}{c}{\textbf{2Wiki}} & \multicolumn{2}{c}{\textbf{Musique}} & \multicolumn{2}{c}{\textbf{Bamboogle}} & \multicolumn{2}{c}{\textbf{Avg.}} \\
\cmidrule(lr){3-4}\cmidrule(lr){5-6}\cmidrule(lr){7-8}\cmidrule(lr){9-10}\cmidrule(lr){11-12}
\textbf{top $k$} & \textbf{Retriever} & EM & F1 & EM & F1 & EM & F1 & EM & F1 & EM & F1 \\
\midrule
\multirow{3}{*}{$k=1$}
 & Stella-400M             & 0.4149 & 0.5284 & 0.4367 & 0.4946 & 0.1849 & 0.2684 & 0.3520 & 0.4963 & 0.3472 & 0.4470 \\
 & Agentic-R & 0.4212 & 0.5328 & 0.4392 & 0.4966 & 0.1833 & 0.2700 & 0.4240 & 0.5260 & 0.3670 & 0.4564 \\
 & \cellcolor{criticblue}\textbf{Critic-Embed}   & \cellcolor{criticblue}\textbf{0.4289} & \cellcolor{criticblue}\textbf{0.5446} & \cellcolor{criticblue}\textbf{0.4497} & \cellcolor{criticblue}\textbf{0.5083} & \cellcolor{criticblue}\textbf{0.1907} & \cellcolor{criticblue}\textbf{0.2821} & \cellcolor{criticblue}\textbf{0.4480} & \cellcolor{criticblue}\textbf{0.5872} & \cellcolor{criticblue}\textbf{0.3794} & \cellcolor{criticblue}\textbf{0.4806} \\
\midrule
\multirow{3}{*}{$k=3$}
 & Stella-400M             & 0.4540 & 0.5698 & 0.4479 & 0.5059 & 0.2085 & 0.2988 & 0.4880 & 0.6213 & 0.3996 & 0.4990 \\
 & Agentic-R & 0.4492 & 0.5659 & 0.4643 & 0.5255 & 0.2127 & 0.3013 & 0.4880 & 0.5959 & 0.4036 & 0.4972 \\
 & \cellcolor{criticblue}\textbf{Critic-Embed}   & \cellcolor{criticblue}\textbf{0.4610} & \cellcolor{criticblue}\textbf{0.5811} & \cellcolor{criticblue}\textbf{0.4700} & \cellcolor{criticblue}\textbf{0.5309} & \cellcolor{criticblue}\textbf{0.2242} & \cellcolor{criticblue}\textbf{0.3185} & \cellcolor{criticblue}\textbf{0.4960} & \cellcolor{criticblue}\textbf{0.6269} & \cellcolor{criticblue}\textbf{0.4128} & \cellcolor{criticblue}\textbf{0.5144} \\
\midrule
\multirow{3}{*}{$k=5$}
 & Stella-400M             & 0.4578 & 0.5789 & 0.4575 & 0.5151 & 0.2321 & 0.3249 & 0.5120 & 0.6285 & 0.4149 & 0.5119 \\
 & Agentic-R & 0.4540 & 0.5743 & 0.4682 & 0.5313 & 0.2238 & 0.3148 & 0.4960 & 0.6211 & 0.4105 & 0.5104 \\
 & \cellcolor{criticblue}\textbf{Critic-Embed}   & \cellcolor{criticblue}\textbf{0.4686} & \cellcolor{criticblue}\textbf{0.5905} & \cellcolor{criticblue}\textbf{0.4763} & \cellcolor{criticblue}\textbf{0.5367} & \cellcolor{criticblue}\textbf{0.2346} & \cellcolor{criticblue}\textbf{0.3277} & \cellcolor{criticblue}\textbf{0.5280} & \cellcolor{criticblue}\textbf{0.6536} & \cellcolor{criticblue}\textbf{0.4269} & \cellcolor{criticblue}\textbf{0.5272} \\
\bottomrule
\end{tabular}
}
\vspace{-0.2cm}
\caption{\textbf{Multi-hop QA --- retriever comparison.} Three dense retrievers evaluated at top $k \in \{1, 3, 5\}$ under the same Search-R1 reasoner. \textbf{Bold} = best in column for that $k$.}
\label{tab:retriever}
\vspace{-0.2cm}
\end{table*}

\subsection{Results}
\paragraph{RQ1: Can inference-time refinement close the retrieval gap on a frozen retriever?}
\label{sec:exp:rq1}
We hold reasoner family (Search-R1), retriever (frozen Stella-400M, top $k{=}1$), and refinement budget ($K{=}2$) fixed, and vary (i) the reasoner scale (3B / 7B / 14B) and (ii) the critic scale (14B / 32B / 72B). Results reported in Table~\ref{tab:its-multihop}.

\textbf{(1) Any critic beats no critic.}
For every (reasoner, dataset, metric) cell, the smallest critic (14B) already strictly improves over the \textit{no-critic} ablation. The lift is large even on the hardest datasets (Bamboogle, Musique), confirming that the gains are driven by the critic's verdict + instruction rewrite rather than by the extra forward passes alone.

\textbf{(2) Inference-time scaling is not strictly monotonic: a larger critic does not guarantee better generation.} Rather than a linear improvement, scaling the critic from 32B to 72B yields sharply diminishing returns and occasional performance degradation on complex tasks. The 72B critic reliably boosts the weaker 3B reasoner across the board, but the 32B critic proves optimal for harder datasets like Musique, and on Bamboogle when paired with the 7B reasoner. Ultimately, averaged across the suite, the 32B critic edges out the 72B for the 7B reasoner (0.3293 / 0.4176 vs. 0.3192 / 0.4108 EM/F1), showing that beyond 32B parameters, injecting more evaluator compute may not overcome the limitations of the reasoner and retriever.

\begin{table*}[ht]
\centering
\resizebox{1.75\columnwidth}{!}{%
\small
\setlength{\tabcolsep}{4pt}
\renewcommand{\arraystretch}{1.1}
\begin{tabular}{@{}l cc cc cc cc cc@{}}
\toprule
& \multicolumn{2}{c}{\textbf{HotpotQA}} & \multicolumn{2}{c}{\textbf{2Wiki}} & \multicolumn{2}{c}{\textbf{Musique}} & \multicolumn{2}{c}{\textbf{Bamboogle}} & \multicolumn{2}{c}{\textbf{Avg.}} \\
\cmidrule(lr){2-3}\cmidrule(lr){4-5}\cmidrule(lr){6-7}\cmidrule(lr){8-9}\cmidrule(lr){10-11}
\textbf{Method} & EM & F1 & EM & F1 & EM & F1 & EM & F1 & EM & F1 \\
\midrule
Search-R1 (static, Stella-400M)            & 0.4149 & 0.5284 & 0.4367 & 0.4946 & 0.1849 & 0.2684 & 0.3520 & 0.4963 & 0.3472 & 0.4470 \\
\rowcolor{criticblue} Critic-Embed (static, no loop)             & 0.4289 & 0.5446 & 0.4497 & 0.5083 & 0.1907 & 0.2821 & 0.4480 & 0.5872 & 0.3794 & 0.4806 \\
\rowcolor{criticblue} Critic-R-Zero (loop on Stella-400M)        & \textbf{0.4425} & \textbf{0.5593} & 0.4580 & 0.5155 & \textbf{0.2127} & \textbf{0.3045} & 0.4480 & 0.5627 & 0.3903 & 0.4855 \\
\rowcolor{criticblue} \textbf{Critic-R} (loop on Critic-Embed)   & 0.4365 & 0.5501 & \textbf{0.4634} & \textbf{0.5237} & 0.2027 & 0.2895 & \textbf{0.4800} & \textbf{0.6200} & \textbf{0.3957} & \textbf{0.4959} \\
\bottomrule
\end{tabular}
}
\vspace{-0.2cm}
\caption{\textbf{Full Critic-R results on Multi-Hop QA.} All configurations share the same Search-R1 (14B) reasoner; the critic (when active) is Qwen2.5-72B; top $k\!=\!1$; $K\!=\!2$ refinement attempts. \textbf{Bold} = best in column.}
\label{tab:critic-embed}
\end{table*}

\begin{table*}[ht]
\centering
\resizebox{1.75\columnwidth}{!}{%
\small
\renewcommand{\arraystretch}{1.1}
\resizebox{\textwidth}{!}{%
\begin{tabular}{@{}ll cc cc cc cc cc@{}}
\toprule
& & \multicolumn{2}{c}{\textbf{HotpotQA}} & \multicolumn{2}{c}{\textbf{2Wiki}} & \multicolumn{2}{c}{\textbf{Musique}} & \multicolumn{2}{c}{\textbf{Bamboogle}} & \multicolumn{2}{c}{\textbf{Avg.}} \\
\cmidrule(lr){3-4}\cmidrule(lr){5-6}\cmidrule(lr){7-8}\cmidrule(lr){9-10}\cmidrule(lr){11-12}
\textbf{top $k$} & \textbf{Method} & EM & F1 & EM & F1 & EM & F1 & EM & F1 & EM & F1 \\
\midrule
\multirow{2}{*}{$k=1$}
 & \cellcolor{criticblue}\textbf{Critic-Embed} & \cellcolor{criticblue}\textbf{0.4289} & \cellcolor{criticblue}\textbf{0.5446} & \cellcolor{criticblue}\textbf{0.4497} & \cellcolor{criticblue}\textbf{0.5083} & \cellcolor{criticblue}\textbf{0.1907} & \cellcolor{criticblue}\textbf{0.2821} & \cellcolor{criticblue}0.4480 & \cellcolor{criticblue}\textbf{0.5872} & \cellcolor{criticblue}\textbf{0.3794} & \cellcolor{criticblue}\textbf{0.4806} \\
 & \quad w/o Introspective Feedback ($T_{i+1}$)           & 0.3932          & 0.5098          & 0.4174          & 0.4749          & 0.1787          & 0.2623          & \textbf{0.4560} & 0.5613 & 0.3614 & 0.4521 \\
\midrule
\multirow{2}{*}{$k=3$}
 & \cellcolor{criticblue}\textbf{Critic-Embed} & \cellcolor{criticblue}\textbf{0.4610} & \cellcolor{criticblue}\textbf{0.5811} & \cellcolor{criticblue}\textbf{0.4700} & \cellcolor{criticblue}\textbf{0.5309} & \cellcolor{criticblue}\textbf{0.2242} & \cellcolor{criticblue}\textbf{0.3185} & \cellcolor{criticblue}\textbf{0.4960} & \cellcolor{criticblue}\textbf{0.6269} & \cellcolor{criticblue}\textbf{0.4128} & \cellcolor{criticblue}\textbf{0.5144} \\
 & \quad w/o Introspective Feedback ($T_{i+1}$)           & 0.4319          & 0.5454          & 0.4474          & 0.5059          & 0.2019          & 0.2924          & 0.4800          & 0.6109          & 0.3903 & 0.4887 \\
\midrule
\multirow{2}{*}{$k=5$}
 & \cellcolor{criticblue}\textbf{Critic-Embed} & \cellcolor{criticblue}\textbf{0.4686} & \cellcolor{criticblue}\textbf{0.5905} & \cellcolor{criticblue}\textbf{0.4763} & \cellcolor{criticblue}\textbf{0.5367} & \cellcolor{criticblue}\textbf{0.2346} & \cellcolor{criticblue}\textbf{0.3277} & \cellcolor{criticblue}\textbf{0.5280} & \cellcolor{criticblue}\textbf{0.6536} & \cellcolor{criticblue}\textbf{0.4269} & \cellcolor{criticblue}\textbf{0.5272} \\
 & \quad w/o Introspective Feedback ($T_{i+1}$)           & 0.4400          & 0.5568          & 0.4536          & 0.5142          & 0.2147          & 0.3052          & 0.4800          & 0.6197          & 0.3971 & 0.4990 \\
\bottomrule
\end{tabular}%
}
}
\vspace{-0.2cm}
\caption{Effect of removing the agent's introspective feedback when collecting training trajectories for Critic-Embed.}
\label{tab:ablation-feedback}
\vspace{-0.2cm}
\end{table*}

\paragraph{RQ2: Do Critic-R-Zero trajectories transfer into a better retriever?}
\label{sec:exp:rq2}
We evaluate Critic-Embed as a drop-in replacement for the retriever, \emph{with no loop active}. This isolates the supervision signal that the trained retriever absorbs from Critic-R-Zero trajectories. Table~\ref{tab:retriever} reports three retrievers with identical agent and conditions: the off-the-shelf Stella-400M backbone, the Agentic-R retriever baseline, and our Critic-Embed. 

\begin{figure}[t]
\centering
\includegraphics[width=0.9\columnwidth]{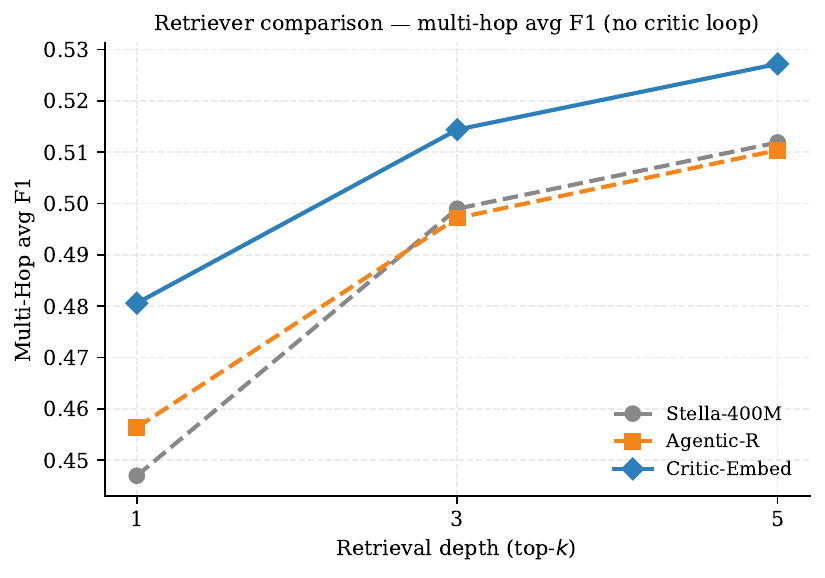}
\vspace{-0.3cm}
\caption{Multi-hop average F1 vs.\ retrieval depth $k$.}
\label{fig:ret-vs-k}
\vspace{-0.2cm}
\end{figure}

\textbf{Critic-Embed is the best-performing retriever in every setting.}
The largest absolute gains come at \textbf{top $k=1$}, where retrieval errors are most costly: on Bamboogle, EM/F1 climbs from 0.3520 / 0.4963 (Stella-400M) and 0.4240 / 0.5260 (Agentic-R) to \textbf{0.4480 / 0.5872} with Critic-Embed; the multi-hop average rises from 0.3472 / 0.4470 to 0.3794 / 0.4806. As $k$ increases the absolute gap narrows because of the higher recall, yet Critic-Embed retains the lead at $k=3$ (average 0.4128 / 0.5144 vs.\ 0.3996 / 0.4990 for Stella-400M and 0.4036 / 0.4972 for Agentic-R) and $k=5$ (average 0.4269 / 0.5272 vs.\ 0.4149 / 0.5119 and 0.4105 / 0.5104). Figure~\ref{fig:ret-vs-k} visualizes the multi-hop average for the three retrievers across $k$. The result establishes that the trajectories produced by Critic-R-Zero contain genuine, transferable retrieval supervision: even before any inference-time criticism is layered on top, training retriever on those trajectories outperforms a retriever that was co-trained end-to-end with an agent on the same task.

\paragraph{RQ3: Does combining the inference-time loop and the trained retriever yield further gains?}
\label{sec:exp:rq3}

Having established that the inference-time query refinement loop and the trained retriever each close part of the retrieval gap on their own, we now ask whether combining them yields further gains. Table~\ref{tab:critic-embed} reports four configurations on the same Search-R1 (14B) reasoner at top $k\!=\!1$: the static \textbf{Search-R1} baseline (Stella-400M backbone, no critic loop); \textbf{Critic-Embed} as a static retriever (no loop); \textbf{Critic-R-Zero} (the loop running on top of the frozen Stella-400M); and the full \textbf{Critic-R} system (the loop running on top of Critic-Embed).

Three observations emerge. (1) Critic-Embed alone, with no inference-time loop, already lifts the multi-hop average over the static Stella-400M baseline from 0.3472 / 0.4470 to 0.3794 / 0.4806 EM/F1, without any test-time refinement. (2) the Critic-R-Zero loop on the frozen backbone reaches 0.3903 / 0.4855, showing that inference-time refinement and retriever fine-tuning each close part of the same overall gap, with the loop modestly ahead on this configuration. (3) combining them, yields the best overall configuration: average \textbf{0.3957 EM / 0.4959 F1}, exceeding both Critic-Embed alone and Critic-R-Zero alone. The per-dataset picture is mixed: Critic-R wins decisively on Bamboogle (0.4800 / 0.6200 vs.\ 0.4480 / 0.5627 for Critic-R-Zero) and on 2Wiki, while Critic-R-Zero edges ahead on HotpotQA and Musique. The two configurations are therefore not redundant: each repairs a different slice of retrieval failures, and the loop and the trained retriever are best read as complementary contributions rather than alternatives.

\paragraph{RQ4: Is the agent's introspective feedback a key source of the supervisory signal?}
\label{sec:exp:rq4}

A central design choice of Critic-R-Zero is that the critic does not judge retrievals from the query and documents alone: it is conditioned on the reasoning agent's own introspective trace $T_{i+1}$ over the retrieved passages. We claim that this conditioning is essential.  We test that directly by re-collecting training trajectories with a modified Critic-R-Zero in which the speculative-feedback step is removed and the critic receives only the global question $Q$, the generated query $q_i$, and the retrieved documents $D_i$. We then fine-tune a separate retriever, denoted ``w/o Introspective Feedback ($T_{i+1}$)'', on these alternative trajectories using the same recipe as Critic-Embed, and evaluate it under the same no-loop static-retriever protocol used in Table~\ref{tab:retriever}.

Table~\ref{tab:ablation-feedback} reports the results. At every retrieval depth, removing the introspective feedback $T_{i+1}$ strictly degrades the resulting retriever. The drop in multi-hop average is substantial \textbf{$-0.0180$} EM and \textbf{$-0.0285$} F1 at $k\!=\!1$, $-0.0225$ EM and $-0.0257$ F1 at $k\!=\!3$, and $-0.0298$ EM and $-0.0282$ F1 at $k\!=\!5$ — and consistent across HotpotQA, 2Wiki, Musique, and (with a small exception on Bamboogle EM at $k\!=\!1$) Bamboogle. This indicates that the agent's introspection contains signals of dissatisfaction of the agent about the retrieved documents and is not just a marginal input to the critic but a primary source of the supervisory signal that Critic-Embed inherits: without it, the critic's verdicts are noisier, the trajectories are weaker, and the distilled retriever inherits the deficit. So, inference-time scaling alone doesn't make agentic search better. The result supports our reading that the critic specializes \emph{over} the reasoner's introspection rather than independently of it.

\section{Conclusion}

In this work, we demonstrate that retriever remains a critical bottleneck in agentic search. We introduced Critic-R, a framework where a dedicated critic evaluates retrieved evidence against the reasoning agent's introspective trace, with two complementary mechanisms: Critic-R-Zero, an inference-time procedure that iteratively refines queries and retrieval instructions, and Critic-Embed, a retriever fine-tuned on the contrastive trajectories produced by this procedure without manual relevance annotation. Evaluated across several challenging multi-hop QA benchmarks, the combined Critic-R system achieved substantial improvements in downstream task accuracy, proving that explicitly modeling and optimizing retrieval quality from within the agentic loop is a powerful path toward more robust agentic search.

\section*{Limitations}

The success of the critic relies heavily on the reasoning agent's capacity to give feedback about the retrieved documents or identify missing information. While state-of-the-art RL-tuned reasoning models (like Search-R1) naturally possess this ability, weaker or smaller language models may struggle to produce accurate introspective traces, thereby degrading the critic's verification signal. Furthermore, our experiments primarily focus on multi-hop and general knowledge-intensive question answering using a static Wikipedia corpus. The behavior and efficacy of the Critic-R has not yet been evaluated in highly dynamic environments, such as real-time web search or private enterprise document systems, where corpus noise and distribution shifts are drastically more pronounced.

\section*{Acknowledgments}
This work was supported in part by the Center for Intelligent Information Retrieval, in part by NSF grant \#2402873, in part by the Office of Naval Research contract \#N000142412612, and with support from Google.org. Any opinions, findings and conclusions or recommendations expressed in this material are those of the authors and do not necessarily reflect those of the sponsor.


\bibliography{custom}

\appendix

\section{General-Domain QA Results}
\label{app:general-qa}

Table~\ref{tab:its-general} reports the inference-time scaling results for \textbf{Critic-R-Zero} on the three general-domain QA benchmarks (NQ, TriviaQA, PopQA), complementing the multi-hop results in Table~\ref{tab:its-multihop}. The same trends observed on the multi-hop suite hold here: any critic reliably improves over the \textit{no-critic} ablation across all reasoner scales, and the largest critic (Qwen2.5-72B) typically yields the strongest average performance.

\begin{table*}[t]
\centering
\small
\setlength{\tabcolsep}{4pt}
\renewcommand{\arraystretch}{1.1}
\begin{tabular}{@{}ll cc cc cc cc@{}}
\toprule
& & \multicolumn{2}{c}{\textbf{NQ}} & \multicolumn{2}{c}{\textbf{TriviaQA}} & \multicolumn{2}{c}{\textbf{PopQA}} & \multicolumn{2}{c}{\textbf{Avg.}} \\
\cmidrule(lr){3-4}\cmidrule(lr){5-6}\cmidrule(lr){7-8}\cmidrule(lr){9-10}
\textbf{Reasoner} & \textbf{Critic ($\mathcal{M}_C$)} & EM & F1 & EM & F1 & EM & F1 & EM & F1 \\
\midrule
\multirow{4}{*}{Search-R1 (14B)}
 & \textit{no-critic}     & 0.4385 & 0.5275 & 0.6554 & 0.7306 & 0.4069 & 0.4444 & 0.5003 & 0.5675 \\
 & Qwen2.5-14B           & 0.4452 & 0.5340 & 0.6619 & 0.7367 & 0.4156 & 0.4528 & 0.5076 & 0.5745 \\
 & Qwen2.5-32B           & \textbf{0.4526} & \textbf{0.5430} & 0.6668 & 0.7424 & 0.4176 & 0.4570 & 0.5124 & 0.5808 \\
 & Qwen2.5-72B           & 0.4499 & 0.5406 & \textbf{0.6687} & \textbf{0.7438} & \textbf{0.4274} & \textbf{0.4667} & \textbf{0.5154} & \textbf{0.5837} \\
\midrule
\multirow{4}{*}{Search-R1 (7B)}
 & \textit{no-critic}     & 0.3582 & 0.4587 & 0.5859 & 0.6662 & 0.3377 & 0.3912 & 0.4273 & 0.5054 \\
 & Qwen2.5-14B           & 0.3731 & 0.4755 & 0.5975 & 0.6794 & 0.3509 & 0.4076 & 0.4405 & 0.5209 \\
 & Qwen2.5-32B           & \textbf{0.3801} & \textbf{0.4823} & 0.6056 & 0.6867 & 0.3549 & 0.4093 & 0.4469 & 0.5261 \\
 & Qwen2.5-72B           & 0.3740 & 0.4806 & \textbf{0.6063} & \textbf{0.6872} & \textbf{0.3657} & \textbf{0.4229} & \textbf{0.4487} & \textbf{0.5303} \\
\midrule
\multirow{4}{*}{Search-R1 (3B)}
 & \textit{no-critic}     & 0.3044 & 0.4048 & 0.5058 & 0.5854 & 0.3031 & 0.3557 & 0.3711 & 0.4487 \\
 & Qwen2.5-14B           & 0.3102 & 0.4107 & 0.5244 & 0.6035 & 0.3205 & 0.3731 & 0.3851 & 0.4625 \\
 & Qwen2.5-32B           & 0.3199 & 0.4164 & \textbf{0.5366} & \textbf{0.6151} & 0.3229 & 0.3755 & 0.3932 & 0.4690 \\
 & Qwen2.5-72B           & \textbf{0.3211} & \textbf{0.4193} & 0.5316 & 0.6122 & \textbf{0.3287} & \textbf{0.3827} & \textbf{0.3938} & \textbf{0.4714} \\
\bottomrule
\end{tabular}
\caption{\textbf{General QA --- inference-time scaling along the critic-size axis} for \textbf{Critic-R-Zero} (frozen Stella-400M retriever, top-$k\!=\!1$, $K\!=\!2$ refinement attempts). \textbf{Bold} marks the best EM/F1 per (reasoner, dataset, metric).}
\label{tab:its-general}
\end{table*}

\section{Implementation Details}
\label{app:impl-details}

\paragraph{Reasoning Agent ($\mathcal{M}_R$):} The reasoning agent is an instruction-tuned LLM operating under the ReAct paradigm \cite{yao2022react}. At each step, the agent first emits a reasoning trace enclosed in \texttt{<think>}\dots\texttt{</think>} tags, followed by an action drawn from two types: a search action \texttt{<search>}$q$\texttt{</search>}, which issues a sub-query $q$ to the retriever when external evidence is required, or a final answer action \texttt{<answer>}$\hat{y}$\texttt{</answer>}, which terminates the trajectory. Retrieved documents returned by the retriever are injected back into the agent's context within \texttt{<information>}\dots\texttt{</information>} tags, and the agent is explicitly instructed to use its subsequent thinking trace to verbalize which aspects of the retrieved evidence are missing or misaligned with its current sub-goal. This introspective feedback is what the critic subsequently exploits to detect retrieval failures. The full system prompt is provided in Figure~\ref{fig:prompt-reasoner} in Appendix~\ref{app:reasoner-prompt}.

\paragraph{Critic Model ($\mathcal{M}_C$):} The critic is a separate LLM, and it operates in two sequential modes that decouple the judgment of retrieval quality from the act of refining it. In the \emph{satisfaction judgment} mode, the critic is prompted with the original question $Q$, the current sub-query $q_i^{(t)}$, the retrieved documents $D_i^{(t)}$, and the reasoner's introspective thinking trace $T_{i+1}^{(t)}$, and is asked to emit a binary verdict $\sigma_i^{(t)} \in \{\texttt{yes}, \texttt{no}\}$ within a \texttt{<satisfactory>} tag together with a concise diagnostic reason $r_i^{(t)}$ within a \texttt{<reason>} tag that states precisely what evidence, if any, is missing. In the \emph{query refinement} mode, the critic is invoked only when the previous verdict is negative; it is then prompted with the failed sub-query $q_i^{(t)}$, the failed instruction $I_i^{(t)}$, and the diagnostic reason $r_i^{(t)}$, and is asked to produce a refined retrieval instruction inside an \texttt{<instruction>} tag and a refined sub-query inside a \texttt{<query>} tag for the next retrieval attempt. Splitting the critic into these two modes prevents premature commitment to refinements when retrieval is in fact adequate, and allows the refinement step to focus entirely on diagnosing and bridging the specific gap identified during judgment. The full satisfaction judgment prompt $P_{\text{J}}$ (Figure~\ref{fig:prompt-judge}) and the query refinement prompt $P_{\text{R}}$ (Figure~\ref{fig:prompt-refine}) are provided in Appendix~\ref{app:critic-prompts}.

\section{Dataset Statistics}
\label{app:dataset-stats}

Table~\ref{tab:dataset-stats} reports the evaluation set sizes for the seven QA datasets used in our experiments. Following \citet{jin2025search}, we evaluate on the dev split when an official test split is not publicly available, and otherwise use the test split. The first four datasets are multi-hop QA benchmarks; the last three are general-domain (predominantly single-hop) QA benchmarks.

\begin{table}[H]
\centering
\small
\setlength{\tabcolsep}{5pt}
\renewcommand{\arraystretch}{1.1}
\begin{tabular}{@{}llr@{}}
\toprule
\textbf{Dataset} & \textbf{Split} & \textbf{\# Examples} \\
\midrule
\multicolumn{3}{l}{\emph{Multi-hop QA}} \\
HotpotQA~\cite{yang2018hotpotqa}          & dev  & 7{,}405  \\
2WikiMultihopQA~\cite{ho2020constructing} & dev  & 12{,}576 \\
MuSiQue~\cite{trivedi2022musique}         & dev  & 2{,}417  \\
Bamboogle~\cite{press2023measuring}       & test & 125      \\
\midrule
\multicolumn{3}{l}{\emph{General-domain QA}} \\
NQ~\cite{kwiatkowski2019natural} & test & 3{,}610  \\
TriviaQA~\cite{joshi2017triviaqa}               & test & 11{,}313 \\
PopQA~\cite{mallen2023not}                      & test & 14{,}267 \\
\bottomrule
\end{tabular}
\caption{Evaluation set sizes for the QA datasets used in our experiments.}
\label{tab:dataset-stats}
\end{table}

\section{Critic-Embed Training Details}
\label{app:training-details}

Critic-Embed is initialized from {Stella-400M} embedding model~\cite{zhang2025jasperstelladistillationsota}\footnote{\url{https://hf.co/NovaSearch/stella_en_400M_v5}} and fine-tuned with InfoNCE (temperature $\tau = 0.02$). The effective batch size is 128 (per-device 32 with 4-step gradient accumulation), trained for 5 epochs at learning rate $2{\times}10^{-5}$, weight decay 0.01, linear warmup over the first 10\% of steps, and gradient clipping at 1.0. Mixed precision (FP16) is used throughout. Natural contrastive pairs are oversampled by a factor of 4 relative to positive-only samples, and up to 3 intra-trajectory hard negatives are retained per query. Hard negatives are combined with in-batch negatives.

\section{Prompts}
\label{sec:appendix}

Throughout this section, we use \promptslot{slot} to denote runtime-substituted variables and \prompttag{tag}\,/\,\promptetag{tag} to denote the structured output markers parsed from the model's response.

\subsection{Reasoning Agent Prompt}
\label{app:reasoner-prompt}

The reasoning agent $\mathcal{M}_R$ is driven by a single user-turn prompt that establishes the ReAct interaction protocol. The full template is shown in Figure~\ref{fig:prompt-reasoner}.

\subsection{Critic Model Prompts}
\label{app:critic-prompts}

The critic $\mathcal{M}_C$ is invoked with two distinct prompts corresponding to its two modes: the satisfaction judgment prompt $P_{\text{J}}$ (Figure~\ref{fig:prompt-judge}) and the query refinement prompt $P_{\text{R}}$ (Figure~\ref{fig:prompt-refine}).

\section{Use of AI Assistants}
We use Claude\footnote{\url{https://claude.ai/}} to improve the presentation of the paper.

\begin{figure*}[ht]
\begin{promptbox}{Reasoning Agent ($\mathcal{M}_R$)}
Answer the given question. You must conduct reasoning inside \prompttag{think} and \promptetag{think} first every time you get new information and give feedback indicating what is missing in the information or what needs to be improved in the query.

After reasoning, if you find you lack some knowledge, you can call a search engine by \prompttag{search}\,\promptslot{query}\,\promptetag{search} and it will return the top searched results between \prompttag{information} and \promptetag{information}. You can search as many times as you want.

If you find no further external knowledge needed, you can directly provide the answer inside \prompttag{answer} and \promptetag{answer}, without detailed illustrations. For example, \prompttag{answer}\,Beijing\,\promptetag{answer}.

\vspace{2pt}\hrule\vspace{4pt}
\textbf{Question:} \promptslot{question}
\end{promptbox}
\caption{System prompt for the reasoning agent $\mathcal{M}_R$. The placeholder \promptslot{question} is replaced at runtime with the input question.}
\label{fig:prompt-reasoner}
\end{figure*}

\begin{figure*}[!t]
\begin{promptbox}{Satisfaction Judgment Prompt $P_{\text{J}}$}
You are an evaluator for a search problem. You will be given a global search query, a local sub-query, retrieved documents, and critique feedback from a reasoning model. Your ONLY task is to evaluate if the retrieved documents are satisfactory for answering the \textbf{current sub-query}.

\vspace{2pt}\hrule\vspace{4pt}
\textbf{Global Query:}~\promptslot{og\_query}\\
\textbf{Local Sub-query:}~\promptslot{sub\_query}\\
\textbf{Retrieved Documents:}~\promptslot{documents}\\
\textbf{Critique of Retrieved Documents:}~\promptslot{feedback}
\vspace{2pt}\hrule\vspace{4pt}

\textbf{Outputs:}
\begin{itemize}\setlength{\itemsep}{2pt}\setlength{\topsep}{2pt}
\item \textbf{satisfactory:} If documents are sufficient to answer the current sub-query, output \prompttag{satisfactory}\,yes\,\promptetag{satisfactory}; otherwise \prompttag{satisfactory}\,no\,\promptetag{satisfactory}.
\item \textbf{reason:} A concise reason for the decision. If `no', state exactly what is missing. Format: \prompttag{reason}\,\promptslot{reason}\,\promptetag{reason}.
\end{itemize}
\end{promptbox}
\caption{Satisfaction judgment prompt $P_{\text{J}}$. Given the global question, the current sub-query, the retrieved documents, and the reasoner's introspective feedback, the critic emits a binary verdict together with a diagnostic reason.}
\label{fig:prompt-judge}
\end{figure*}

\begin{figure*}[!t]
\begin{promptbox}{Query Refinement Prompt $P_{\text{R}}$}
You are a search query optimizer. The previous search failed to retrieve satisfactory documents.

\vspace{2pt}\hrule\vspace{4pt}
\textbf{Global Query:}~\promptslot{og\_query}\\
\textbf{Failed Sub-query:}~\promptslot{sub\_query}\\
\textbf{Failed Instruction:}~\promptslot{instruction}\\
\textbf{Reason for Failure:}~\promptslot{reason}
\vspace{2pt}\hrule\vspace{4pt}

Based on the Reason for Failure, generate a refined instruction and a refined sub-query to successfully retrieve the missing information.

\textbf{Outputs:}
\begin{itemize}\setlength{\itemsep}{2pt}\setlength{\topsep}{2pt}
\item \textbf{new instruction:} A concise, refined instruction for the next retrieval attempt. Format: \prompttag{instruction}\,\promptslot{refined instruction}\,\promptetag{instruction}.
\item \textbf{new query:} A concise, refined query. Format: \prompttag{query}\,\promptslot{refined query}\,\promptetag{query}.
\end{itemize}
\end{promptbox}
\caption{Query refinement prompt $P_{\text{R}}$. Invoked only when $P_{\text{J}}$ returns \texttt{no}, the critic uses the diagnostic reason to rewrite the failed sub-query and retrieval instruction for the next attempt.}
\label{fig:prompt-refine}
\end{figure*}

\end{document}